\documentclass{raa}            

\usepackage{graphicx,times}             
\graphicspath{{fig/}}
\usepackage{natbib}
\usepackage{color}
\usepackage{amssymb,amsmath}
\usepackage{hyperref}
\hypersetup{pdftitle =paper, pdfauthor =G Li, pdfsubject=SEPs, pdfkeywords = SEP CME }
\hypersetup{colorlinks = true, linkcolor = blue, anchorcolor = red, citecolor = blue, filecolor = red, urlcolor = blue}

\newcommand{\beq}{\begin{equation}}
\newcommand{\eeq}{\end{equation}}
\newcommand{\beqa}{\begin{eqnarray}}
\newcommand{\eeqa}{\end{eqnarray}}

\begin{document}

\title{Probing shock geometry via the charge to mass ratio dependence of heavy ion spectra from 
multiple spacecraft observations of the 2013 November 4 event}

   \volnopage{Vol.0 (200x) No.0, 000--000}      
   \setcounter{page}{1}          

\author{Lulu Zhao \inst{1}, Gang Li\inst{2,*}, 
G. M. Mason\inst{3}, C. Cohen\inst{4}, R. A. Mewaldt\inst{4}, 
M. I. Desai\inst{5,6}, R. W. Ebert\inst{5}, M. A. Dayeh\inst{5}}

\institute{
{Department of Physics and Space Sciences, Florida Institute of Technology, FL, USA. } \\
{Department of Space Science and CSPAR,  gang.li@uah.edu,
University of Alabama in Huntsville, Huntsville, Alabama, USA.} \\
{Applied Physics Laboratory, Johns Hopkins University, Laurel, MD 20723, USA} \\
{California Institute of Technology, Pasadena, CA, 91125, USA} \\
{Southwest Research Institute, San Antonio, TX, USA} \\
{University of Texas at San Antonio, San Antonio, TX, USA} }

   \date{Received~~2016 month day; accepted~~2015~~month day}


\abstract{In large SEP events, ions can be accelerated at CME-driven shocks to very high energies.
Spectra of heavy ions in many large SEP events show features such as roll-overs or spectral breaks.
In some events when the spectra are plotted in energy/nucleon they can be shifted relative to each other to make 
the spectral breaks align. 
The amount of shift is charge-to-mass ratio (Q/A) dependent and varies from event to event.
This can be understood if the spectra of heavy ions are organized by the 
diffusion coefficients \citep{Cohen2005}.
In the work of \citet{Li2009}, the Q/A dependences of the scaling is related to shock geometry 
when the CME-driven shock is close to the Sun. 
For events where multiple in-situ spacecraft observations exist, 
one may expect that different spacecraft are connected to different portions of the CME-driven
 shock that have different shock geometries, therefore yielding different Q/A dependence.
In this work, we examine one SEP event which occurred on 2013 November 4. 
We study the Q/A dependence of the energy scaling for heavy ion spectra using Helium, oxygen and iron ions.
Observations from STEREO-A, STEREO-B and ACE are examined. We find that the scalings are different for 
different spacecraft. We suggest that this is because ACE, STEREO-A and STEREO-B are connected to different 
parts of the shock that have different shock geometries. 
Our analysis indicates that studying the Q/A scaling of in-situ particle spectra can serve as
 a powerful tool to remotely examine the shock geometry for large SEP events.}

   \authorrunning{Zhao, Li, \& Mason et al. }            
   \titlerunning{Shock geometry from multiple spacecraft observation}         

   \maketitle

%
%
\section{Introduction}           
\label{sec.intro}
Understanding Solar Energetic Particles (SEPs) is a central topic of Space Plasma research. 
Studying SEPs provides a unique opportunity to examine the underlying particle acceleration process which 
exists at a variety of astrophysical sites. Furthermore, understanding SEPs is of practical importance since
 SEPs are a major concern of space weather.  It is now widely accepted that these high energy particles are 
accelerated mostly at solar flares and shocks driven by coronal mass ejections (CMEs). 
Events where particles are accelerated mainly at flares are termed ``impulsive'' \citep{Cane1986} as the time 
intensity profile shows a rapid rise and fast decay. In contrast, events where particles are accelerated mainly at CME-driven
shocks are termed ``gradual'' \citep{Cane1986,Reames1999} where the time intensity profiles vary gradually compared to 
impulsive events. For large SEP events, recent studies \citep[e.g.][]{Reames2009, Cliver2006, Gopalswamy2012, Mewaldt2012} 
suggested that energetic particles that are observed near Earth
in these events are mostly accelerated at the shocks driven by CMEs rather than in flare 
active regions.

In many large SEP events, particle fluence spectra exhibit exponential rollover or double power law features
\citep[e.g.][]{Mewaldt2005, Mewaldt2012}. The break energy  or the roll-over energy, $E_0$, is between a few to a 
few 10's of MeV/nucleon \citep{Mazur1992, Cohen2005, Mewaldt2005, Tylka2005, Desai2016}.
Simulations \citep{Li2005} show that spectral breaks can occur naturally for particle acceleration at a CME-driven shock.
In examining these features, \citet{Cohen2003, Cohen2005} and \citet{Mewaldt2005} noted that the break energies are nicely 
ordered by $(Q/A)^{\sigma}$. They suggested that this ordering 
can be understood if the energy breaks or roll-overs for different heavy ions occur at the same values of the diffusion 
coefficient $\kappa$. Later, \citet{Li2009} attempted to relate $\sigma$ to shock geometry. 
They showed that the value of $\sigma$ is usually in the range of $1$ to $2$ for parallel shocks, but can become 
as small as $\sim 1/5$ for perpendicular shocks. 

For the most general case of an oblique shock, the total diffusion coefficient is given by,
\beq
\kappa = \kappa_{||}  \cos^2(\theta_{BN}) + \kappa_{\perp}  \sin^2(\theta_{BN}).
\label{eq:kappa_total}
\eeq
In the above, $\kappa_{||}$ and $\kappa_{\perp}$ are the parallel and perpendicular diffusion coefficients and 
 $\theta_{BN}$ is the angle between the upstream magnetic field and the shock normal. 
Since in general $\kappa_{||}$ and $\kappa_{\perp}$ have different $Q/A$ dependence \citep{Li2009}, 
equation~(\ref{eq:kappa_total}) yields a complicated $Q/A$ dependence for the break energy at an oblique shock. 
Recently, \citet{Desai2016} have surveyed $0.1$-$100$ MeV/nucleon H-Fe fluence spectra for $46$ isolated
large gradual SEP events observed at ACE during solar cycles 23 and 24. They found that the range of  
$\sigma$ for heavy ion spectra in these events is mostly between $0.2$ to $2$, although some events have a 
$\sigma$ value larger than $2$.  

In the work of \citet{Li2009}, it is assumed that the spectral break or roll-over from 
in-situ observations reflects the same feature of the escaped particle spectra at the shock. 
We note that some recent calculations have suggested that spectral breaks can emerge as a transport effect 
\citep{Li2015, Zhao2016}. 
However, in these calculations, the size of the spectral index change,
i.e. $\delta \gamma = \gamma_{a} - \gamma_{b}$, 
where  $\gamma_{a}$ and $\gamma_{b}$ are the spectral indices above and below the break energy $E_0$,
is very small.
This is in contrast to the observations where  $\delta \gamma$ can be large and varies noticeably
from one event to another. Furthermore,  the transport effect shown in \citep{Li2015, Zhao2016}
predicts a $Q/A^{\sigma}$ dependence of the spectra break energy $E_0$ with an upper limit of $\sigma$ to be $1.3$.
In a recent statistical suryey however, \citet{Desai2016b} found that $\sigma$ in $33$ SEP events ranged between 
$\sim 0.2-3$, which clearly exceeds the upper limit of $\sim 1.3$ predicted 
by scatter-dominated transport models \citep{Li2015, Zhao2016}.

Here we follow \citet{Li2009} and assume that the break is a feature of the escaped particle spectrum
at the shock. Note that the Diffusive Shock Acceleration (DSA) does not predict a spectral break for the shock-accelerated particle spectrum.
Nevertheless, there could be a variety of reasons for such a break. For example, 
the break may represent the maximum energy given a finite acceleration time. In this case we expect the
break energies to be high, $>$ several $10$'s of MeVs for protons. It could also represent the cut-off energy for
escape, i.e., particles with energy lower than the break energy are trapped more within the
shock complex. In this case, the break energies may be low, $\sim <$ several MeV for protons. In both cases,
however, the break energy is decided by the diffusion coefficient $\kappa$, so that the same $Q/A$ analysis
discussed in the work of \citet{Li2009} applies.

Here we do not discuss the underlying mechanism that leads to the spectral break, but 
use the $Q/A$ scaling of heavy ion spectra to remotely infer the shock geometry. We note that since
particles are continuously accelerated at the CME-driven shock, this shock geometry reflects only
an ensemble average of the shock geometry over a period. If however, the energetic particles near
the break energy are mostly accelerated at early times (i.e. in the case
of the break energy representing the maximum energy),
then we expect that the spectral break reflects the shock geometry when the shock is still close to the Sun.

\begin{figure}[htb]
    \centering
    \noindent \includegraphics[width=0.6\textwidth]{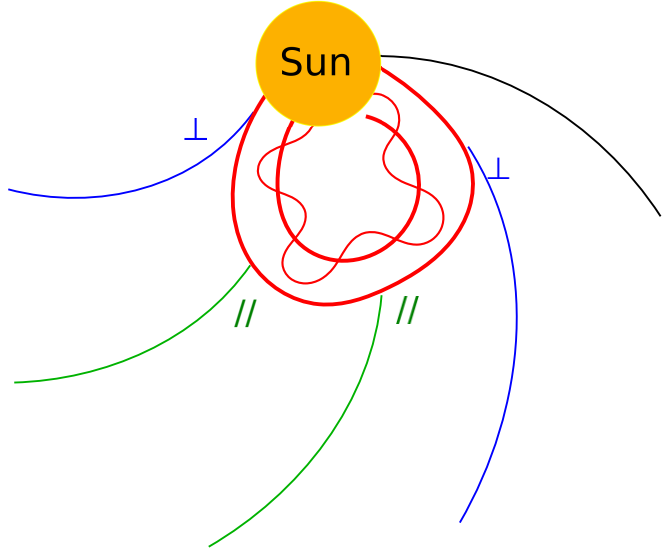}
    \caption{A cartoon showing that two spacecraft can be magnetically connected to different portions of 
a CME-driven shock which have different shock geometries. }
    \label{fig.Cartoon}
\end{figure}

Since for the same CME the shock geometry differs at different longitudes, then with  
simultaneous in-situ observations from multiple spacecraft, one may obtain different 
Q/A-scaling. This is illustrated in the cartoon shown in Figure~\ref{fig.Cartoon}. 
In the cartoon, the two field lines colored in blue (assumed here to be unperturbed Parker field) 
intersect with the shock at a quasi-perpendicular configuration and the two 
field lines colored in green intersect with the shock at a quasi-parallel configuration. 

The above discussion indicates that studying the Q/A scaling of heavy ion spectra simultaneously at multiple 
spacecraft may be used to infer shock geometry of the CME-driven shock for SEP events where heavy ion spectra are 
well organized by Q/A.  In this work, we examine one such SEP event that occurred on 2013 November 4. 
In the following, we describe the observation in section~\ref{sec.obs} and present the fitting results in
section~\ref{sec.res}. We conclude in section~\ref{sec.disc}.

\section{Observations}

\label{sec.obs}
\begin{figure}[htb]
    \centering
    \noindent \includegraphics[width=0.7\textwidth]{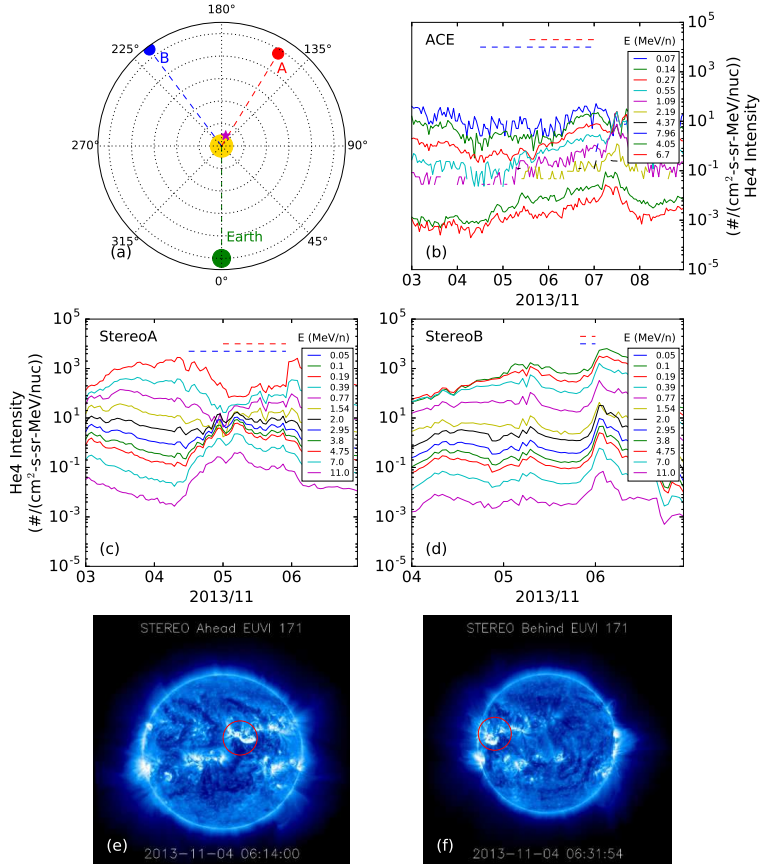}
    \caption{2013 November 4 SEP event observed by three spacecraft. (a) locations of STA, Earth and STB at 2013 November 4 00:00 UT. 
    (b) time intensity profiles of $^{4}$He observed by ACE from 2013 November 3 to November 8. 
    (c) time intensity profiles of $^{4}$He observed by STA from 2013 November 3 to November 6.
    (d) time intensity profiles of $^{4}$He observed by STB from 2013 November 4 to November 6.
    (e) and (f) show the EUVI 171 observations of STA and STB. The active region is marked by the red circle.}
    \label{fig.fig2_obs}
\end{figure}

We study the SEP event using energetic particle measurements obtained by
 the  Ultra-Low Energy Isotope Spectrometer (ULEIS) \citep{Mason1998} and Solar Isotope Spectrometer (SIS) \citep{Stone1998} on the Advanced Composition Explorer (ACE); and
 the Suprathermal Ion Telescope (SIT) \citep{Mason2008} and Low Energy Telescope
 (LET) \citep{Mewaldt2008} on the twin spacecraft Solar TErrestrial RElations
 Observatory (STEREO) A and B.
On 2013 November 4 00:00 UT, the angle between ACE and STEREO-A (STA) was $148.56^\circ$, and the angle between ACE and STEREO-B (STB) was $143.24^\circ$. 
Panel (a) of Figure~\ref{fig.fig2_obs} shows the configuration of STA, STB and ACE for the event and
 the time intensity profiles of Helium for all three spacecraft. 
The eruption occurred on 2013 November 4 05:12:05 UT as identified from the CME catalog
\footnote{\url{http://cdaw.gsfc.nasa.gov/CME\_list/UNIVERSAL/2013\_11/univ2013\_11.html}}. The event is 
also included in the survey of \citet{Richardson2014}. 
The event is a backside halo event as viewed from the Earth; a frontside and slightly western event as viewed from STA; 
and an eastern event from STB. 
Without X-ray observations, we do not know the flare class for this event.
Panels e) and f) show the EUVI 171 observations from STA and STB. 
The active region (AR) is marked by the red circle.

Panels b), c) and d) of Figure~\ref{fig.fig2_obs}
 show the time intensity profiles of helium as observed by ACE, STA, and STB, respectively. 
The event was clearly seen at STA and STB. For ACE, even though it was a backside event,  
one can still see the gradual increase from the background. 
Note that another event from the same AR had occurred on 2013 November 2 (see \citet{Richardson2014}), 
and the SEPs from that event elevated the intensities at all three spacecraft.

The periods we choose for spectral analysis are shown by the dashed lines in panels b), c) and d).  
For STA and ACE, the elevated pre-event background makes it difficult to identify the onset times for 
the lower energy ions. We therefore use two different periods for different energies:
the red dashed lines (November 5 00:00 UT to November 5 22:00 UT) indicate the time interval for the SIT instrument 
and the blue dashed lines (November 4 12:00 UT to November 5 22:00 UT) indicate the time interval for the LET instrument. 
Similarly for ACE we also use two different periods for different energies:
the red dashed lines (November 5 14:00 UT to November 7 02:00 UT) indicates the time interval for the ULEIS instrument and 
the blue dashed lines (November 4 12:00 UT to November 7 02:00 UT) indicate the time interval for the SIS instrument. 
For STB, since this is an eastern event, clear increases of the time-intensity profiles  
do not occur until the end of November 5. So we choose November 5 20:00 UT to November 6 00:00 UT for all energy channels for STB. 
For STA and STB observations,  we choose the stop time of the interval 
as the time at which the intensities peak. 
We do not include energetic particles in the downstream. One reason for doing so is that as shown in \citet{Zank2015}, 
particles can be accelerated at magnetic islands downstream of a shock, leading to an extra acceleration in addition to
 the shock acceleration. This  acceleration may have a different Q/A dependence. Furthermore, turbulence is often stronger 
downstream of a CME-driven shock and additional second-order Fermi acceleration may occur. Such an acceleration is 
also  Q/A dependent. We therefore do not include downstream periods in our analysis. For the ACE
observation, there was no local shock arrival since it was a backside event. We choose the stop time as November 7 00:00 UT, 
which is the onset time of a following event.

In obtaining the integrated energy spectra, we only include energy channels that show a clear increase from the background in the 
time intensity profiles. We do not subtract the pre-event background since the intensities were decaying from a previous event and
the identification of a proper pre-event background is difficult.  
This should not introduce a large uncertainty since the pre-event backgrounds
were well below the intensity levels during the chosen time periods.

The pre-event background does affect the time periods we select for obtaining the spectra.
As we mentioned above, we use different time periods for STA/SIT and STA/LET observations. 
This is because, as shown in panel (c) of Figure~\ref{fig.fig2_obs}, the intensities of lower energy Helium 
(below 2 MeV/nucleon) between November 4 12:00 UT and November 5 00:00 UT 
are still in the decay phase of the previous event. 
If we assume the intensities in these 
energy channels behave similarly to that of the $\sim 2-4$ MeV/nucleon channel, then using the $2-4$ MeV/nucleon intensity 
profile as a reference, we can normalize the time integrated 
intensities for the period denoted by the red dashed line to that denoted by the blue dashed line.
In practice, by noting that for Helium and oxygen observations there is overlap between the LET 
energy channel of $4.25$ MeV/nucleon and the SIT energy channel of $4.37$ MeV/nucleon,
we normalize the intensities of all LET energy channels by multiplying a factor ($1.8$) such that
 the integrated intensity of the LET  $4.25$ MeV/nucleon equals that of the SIT energy channel of $4.37$ MeV/nucleon.
This accounts for the different time intervals used for LET and SIT. 
We follow the same procedure for calculating the ACE spectra and use the energy channel 
of $4.37$ MeV/nucleon of ULEIS and $4.05$ MeV/nucleon of SIS for the normalization. 

\section{Results}
We scale Fe and He spectra to match that of O. The scaling can be seen from the following condition: 
\begin{equation}\label{equ.scale}
 \frac{(Q/A)^{\sigma}_{i}}{ (Q/A)^{\sigma}_{O}} = \frac{E_{i}}{E_{O}}
\end{equation}
where $i$ is He or Fe.

\label{sec.res}
\begin{figure}[htb]
    \centering
    \noindent \includegraphics[width=0.98\textwidth]{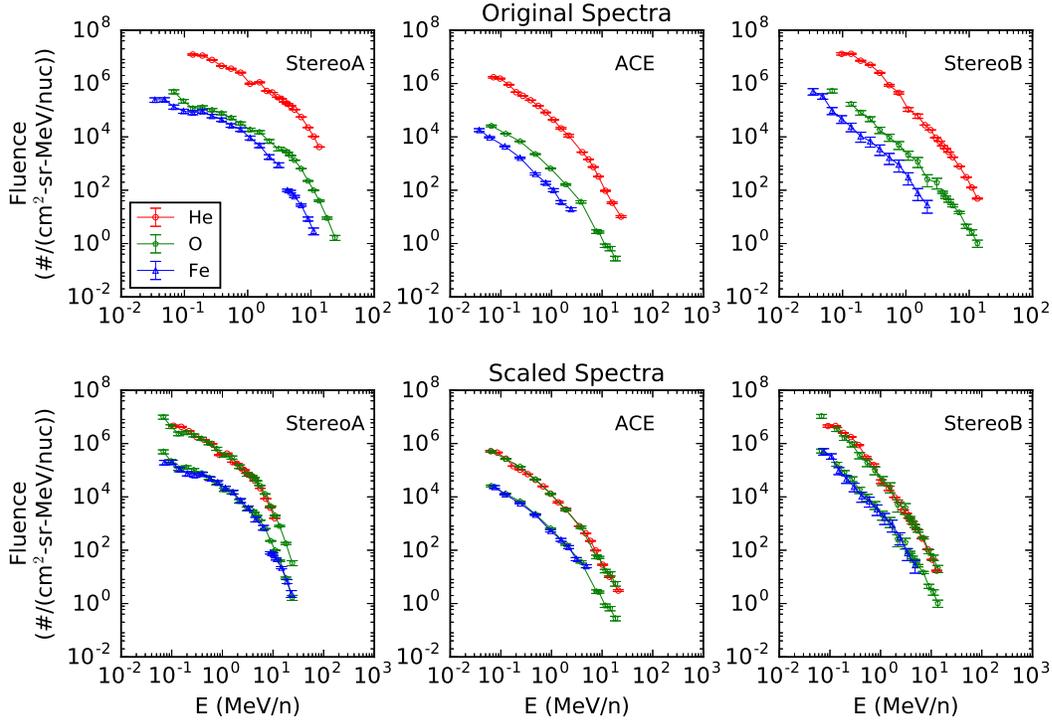}
    \caption{Spectra of Helium, oxygen and iron from STA, ACE, and STB, from the left to the right respectively.
      Upper panels are for the original spectra and lower panels are the scaled spectra. 
      Energy spectra are shifted horizontally and vertically to 
make the comparison easier. }
    \label{fig.SpectraFig3}
\end{figure}

Figure~\ref{fig.SpectraFig3} shows the original and the scaled time-integrated spectra. 
Statistical uncertainties are also shown. 
The upper panels are the original spectra and the lower panels are the spectra after scaling.
In the scaled spectra, for better comparison, 
oxygen spectra are plotted twice, once shifted upward by a factor of $20$. 
The Fe spectra are shifted to the right and the He spectra are shifted to the left to match the ``roll-over'' features of 
the O spectra.  The spectra of Fe and He are also shifted vertically to make the comparison easier.
Statistical uncertainties are shown in the figure.
In the following, however, the scaling factors only  
refer to the energy scaling (i.e. the horizontal shift).
For STA observations, the energy scaling factor for iron is $2.1$, for Helium it is $0.80$.
For STB observations, the energy scaling factor for iron is $1.6$, for Helium it is $0.95$.
For ACE observations, the energy scaling factor for iron is $2.0$, for Helium it is $0.88$.

\begin{figure}[htb]
    \centering
    \noindent \includegraphics[width=0.98\textwidth]{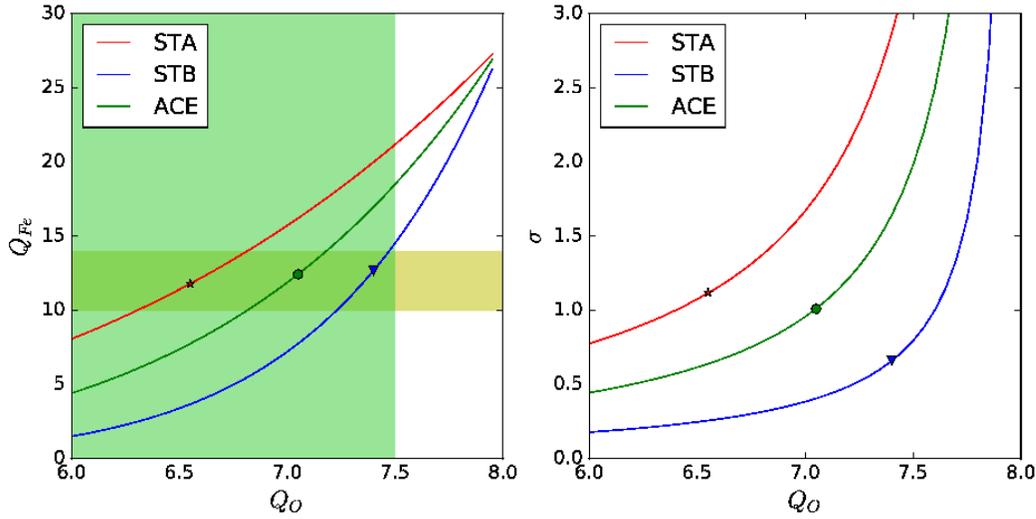}
    \caption{Left: charge state of Fe versus charge state of O. The shaded area represents the most probable 
charge state of O from $6.0$ to $7.5$ and of Fe from $10$ to $14$. Right: coefficient $\sigma$ versus 
charge state of O.}
    \label{fig.Qplot}
\end{figure}

Assuming the charge state of Helium $Q_{He} = 2$, we examine possible charge state of oxygen in the range of
 $Q_O= 6$ to $7.9$. We increase $Q_O$ from $6$ with a step of $\delta Q = 0.1$. 
For each $Q_{O}$ we obtain the corresponding $\sigma$ using equation~(\ref{equ.scale}) with $i=He$ and then using 
equation~(\ref{equ.scale}) again to obtain the charge state of iron $Q_{Fe}$ with $i=Fe$. 
The left panel of Figure~\ref{fig.Qplot} shows the value of 
$Q_{Fe}$ and $Q_{O}$ from our fitting. The red curve is for the STA observations, the blue curve is the STB 
observations and the green curve is for the ACE observations. The shaded area in the left panel represents 
the most probable charge states of oxygen from $6.0$ to $7.5$ and of iron from $10$ to $14$ (see discussions below). 
The right panel shows the corresponding $\sigma$ value. 

For STA (red curve), a charge state of oxygen from $6.0$ to $7.9$ yields a charge state of Fe 
from $8.1$ to $26.5$ and the corresponding $\sigma$ is from $0.8$ to $17.7$.  
For ACE (green curve), a charge state of oxygen from $6.0$ to $7.9$ yields a charge state of $4.4$ to $25.8$ for 
Fe and a corresponding $\sigma$ from $0.44$ to $10.2$. For STB (dark blue curve), 
a charge state of oxygen from $6.0$ to $7.9$ yields a charge state 
of $1.5$ to $24.6$ for Fe and a corresponding $\sigma$ from $0.18$ to $4.1$. 
Note that the range of $\sigma$ shown in Figure~\ref{fig.Qplot} is from $0$ to $3$, similar to that
obtained in \citet{Desai2016}.

The charge states of O and Fe considered by \citet{Cohen2005} are $6.8$ and $11.6$, respectively.
While the charge state for O can be from $6$ to $8$ for any given SEP event, that for Fe is mostly 
between $10.0$ to $14.0$ \citep[e.g.][]{Labrador2005,Mason2012}. If we vary the charge state of iron $Q_{Fe}$ from $10.0$ to $14.0$, then
from Equation~\ref{equ.scale} we find a charge state $Q_{O}=7.23$ to $7.47$, and $\sigma$ = $0.51$ to $0.74$ for STB;
$Q_{O}=6.88$ to $7.22$, and $\sigma$ = $0.84$ to $1.24$ for ACE;
$Q_{O}=6.31$ to $6.80$, and $\sigma$ = $0.94$ to $1.78$ for STA.

In \citet{Tylka2006}, the authors argue that the injection energy increases with shock obliquity, leading to 
a higher $Fe/O$ ratio and higher charge states at quasi-perpendicular shocks. These higher charge states may be from 
previous impulsive SEP material, which typically has greater energies than solar wind material \citep{Tylka2006}. 
\citet{Li2009} obtained a $\sigma \sim 0.22 < 1$ for a perpendicular shock.
In our event, the STB observation yields a $\sigma < 1$ and may correspond to a quasi-perpendicular portion of the shock. 
The STA observation has a $\sigma$ larger than $1$ for most charge states of iron we considered, 
and is consistent with a quasi-parallel shock. The ACE observations lie in between STB and STA.  
If we assume the charge states of both iron and oxygen increase with shock obliquity, and 
that STA (STB) connects to the quasi-parallel (quasi-perpendicular) part of the shock (while ACE connects to a portion of the shock 
having an obliquity in between STA's and STB's),  then a possible choice of 
($Q_{Fe}$, $Q_{O}$, $\sigma$) can be 
($12.7$, $7.4$, $0.66$) for STB, ($12.4$, $7.0$, $1.01$) for ACE, and ($11.8$, $6.6$, $1.12$) for STA.
We choose these values such that the values of $\sigma$ increase from STB to ACE and to STA.  
These choices are labeled as ``star'', ``circle'' and ``triangle'' symbols in Figure~\ref{fig.Qplot}.
This is consistent with the cartoon shown in Figure~\ref{fig.Cartoon} since STB saw an eastern event and STA saw a 
central event. 
For ACE, it is a backside event, the corresponding shock geometry is not clear from Figure~\ref{fig.Cartoon}. 
Comparing the charge state choices for STA and STB, we see that for both oxygen and iron, the charge states of STB observation 
is about $1$ unit larger than those of STA observations. This difference may reflect the injection and seed population dependence on shock 
geometry. We remark that direct charge state measurements for energetic particles from future missions such as 
IMAP will be helpful in resolving this dependence.

\section{Discussion and Conclusion} 
\label{sec.disc}

In this paper, we examine the spectra of heavy ions of the 2013 November 4 SEP event from STA, STB, and ACE. We select this event 
because the time-integrated heavy ion spectra from all three spacecraft show spectral break features above $\sim 1$ Mev/nucleon. 
Although the 
pre-event background is elevated due to the presence of SEPs from another event that had occurred two days earlier, 
time intensity profiles from all
three spacecraft show clear increases from the background.  Our analyses show that: 1) for all three spacecraft the He, O, and Fe
 spectra can be well 
 organized by Q/A and when scaled by $(Q/A)^{\sigma}$, spectra for different heavy ions overlap nicely, 2) the scaling parameter
 $\sigma$ is 
sensitive to the charge to mass ratio,  and for the sets of heavy ion charge states we choose in section~\ref{sec.res}, we obtain 
$\sigma$ for STA, ACE, and STB to be $1.12$, $1.01$, and $0.66$, respectively;
3) Under the framework of \citet{Li2009},  these values of $\sigma$ suggest that 
STA (and ACE) are connected to the quasi-parallel part of the shock and STB is connected to the quasi-perpendicular part of the shock.
This is qualitatively in agreement to the configuration shown in Cartoon~\ref{fig.Cartoon}.

For ACE and STEREO-A, we integrated reasonably long periods (see Figure 2) to obtain the
fluence spectra. As the CME-driven
shock propagates out from the Sun to 1 AU, it continues to accelerate particles, but the maximum energy
decreases with distance. As a result, the event-integrated spectrum represents an ensemble average
of the shock spectra at different times. Since the spectral break feature is around $10$ MeV, a reasonably
high energy for this event, we expect the dominant contributing particles for ACE and STA spectra
are accelerated close to the Sun, e.g, within $0.3$ AU. For STEREO-B, the period of integration is shorter and close
to the shock arrival. Particles of all energies show significant increases from the background at around the same time,
indicating that this is due to magnetic connection.
This is consistent with the fact that the event is an eastern event as seen from STEREO-B. For STEREO-B observations,
one may speculate that the spectrum is more local, i.e. the spectrum represents the shock spectrum when the shock is close
to 1 AU. However, particles accelerated at earlier times but are trapped by the CME-driven shock may
also contribute. Consequently for the STEREO-B observervation, the range of the radial distance of the
shock it samples should be larger
than those by ACE and STEREO-A. Therefore, when we interpret the result of the Q/A scaling, one needs to be
careful in that the shock geometry for the STEREO-B observation may suffer a larger variation than those from STEREO-A and
ACE.

We remark that, as revealed by Figure~\ref{fig.Qplot}, our analysis may be used not only to obtain shock geometry estimates
for multiple spacecraft, but also to examine charge state variability 
of heavy ions in an event. In principle, one can compare our charge state results with that of ion charge state
measurements. However, ACE/SEPICA, which measured energetic particle charge states, does not have data after 2005; and 
both ACE/SWICS and STEREO/PLASTIC make only charge state measurements at solar wind energies.

\begin{acknowledgements}
This work is supported at UAH by NSF grants  AGS-1135432 and AGS-1622391, NASA grant NNX15AJ93G;
at APL by NASA grant NNX13AR20G/115828 (ACE/ULEIS and STEREO/SIT) and NASA subcontract SA4889-26309 
from the University of California Berkeley; at Caltech by  NNX13A66G, NNX11A075G,
and subcontract 00008864 of NNX15AG09G and by NSF grant
AGS-1156004; at SwRI partially by NSF grant AGS-1460118.
\end{acknowledgements}

\label{lastpage}

\end{document}